\documentclass[
conference, 
letterpaper, %
twocolumn,
10pt,
]
{IEEEtran}

\usepackage{mymathmacros}
\usepackage{subfig}
\usepackage[nolist]{acronym}
\usepackage{hyperref}
\usepackage{cleveref}
\usepackage{tikz}
\usepackage{pgfplots}
\usetikzlibrary{decorations.markings}
\usetikzlibrary{shapes}
\usepackage{amsmath}
\usepackage{amssymb}
\usepackage{multirow}
\usepackage[font=small, labelfont=bf]{caption}

\crefname{equation}{}{}

\IEEEoverridecommandlockouts 

\title{A Fully-Unrolled LDPC Decoder Based on Quantized Message Passing} 

\author{
    \IEEEauthorblockN{
       Alexios Balatsoukas-Stimming\IEEEauthorrefmark{1}, 
       Michael Meidlinger\IEEEauthorrefmark{2}, 
       Reza Ghanaatian\IEEEauthorrefmark{1},  
       Gerald Matz\IEEEauthorrefmark{2}, and
       Andreas Burg\IEEEauthorrefmark{1}
    }
    \\
      \begin{tabular}[t]{c@{\extracolsep{1.5cm}}c} 
      \IEEEauthorrefmark{1}EPFL, Switzerland  & \IEEEauthorrefmark{2}Vienna University of Technology, Austria \\
        \small
         Email: \{\href{mailto:alexios.balatsoukas@epfl.ch}{alexios.balatsoukas}, 
                  \href{mailto:reza.ghanaatian@epfl.ch}{reza.ghanaatian},
                  \href{mailto:andreas.burg@epfl.ch}{andreas.burg}\}@epfl.ch
        &
        \small
        Email: \{\href{mailto:mmeidlin@nt.tuwien.ac.at}{mmeidlin},
                 \href{mailto:gmatz@nt.tuwien.ac.at}{gmatz}\}@nt.tuwien.ac.at 
        \\ 
   \end{tabular}\thanks{Funded by WWTF Grant ICT12-054.}
}

\begin{document}

\begin{acronym}
   \acro{MP}{message passing}
   \acro{PMF}{probability mass function}
   \acro{PDF}{probability density function}
   \acro{BICM}{bit-interleaved coded modulation}
   \acro{BI-AWGN}{binary input additive white Gaussian noise}
   \acro{MS}{min-sum}
   \acro{SP}{sum-product}
   \acro{SNR}{signal to noise ratio}
   \acro{BSC}{binary symmetric channel}
   \acro{VN}{variable node}
   \acro{CN}{check node}
   \acro{BER}{bit error rate}
   \acro{BEP}{bit error probability}
   \acro{FER}{frame error rate}
   \acro{FAID}{finite alphabet iterative decoder}
   \acro{IBM}{information bottleneck method}
   \acro{DMC}{discrete memoryless channel}
   \acro{LLR}{log-likelihood ratio}
   \acro{LDPC}{low-density parity-check}
   \acro{LUT}{look-up table}
   \acro{MAP}{maximum a-posteriori probability}
   \acro{MI}{mutual information}
\end{acronym}

\maketitle
\begin{abstract}
In this paper, we propose a finite alphabet message passing algorithm for LDPC codes that replaces the standard min-sum variable node update rule by a mapping based on generic look-up tables. This mapping is designed in a way that maximizes the mutual information between the decoder messages and the codeword bits. We show that our decoder can deliver the same error rate performance as the conventional decoder with a much smaller message bit-width. Finally, we use the proposed algorithm to design a fully unrolled LDPC decoder hardware architecture.
\end{abstract}

\section{Introduction}\label{sec:introduction}

The excellent error correction performance of \ac{LDPC} codes, alongside with the availability of low-complexity and highly 
parallel decoding algorithms and hardware architectures makes them an attractive choice for many high throughput communication systems. 
\ac{LDPC} codes are traditionally decoded using iterative \ac{MP} algorithms like the \ac{SP} algorithm and variants thereof~\cite{chen2005a}, most notably the \ac{MS} algorithm. 
Those conventional algorithms rely on the exchange of continuous messages, which are usually quantized with resolutions of $4$ to $7$ bits in most in hardware implementations. Lower resolutions are possible but entail severe performance penalties, especially in the error-floor region \cite{zhang2009a}. 

Previous work on quantized \ac{MP} algorithms for \ac{LDPC} decoding has shown that decoders which are designed to operate directly on message 
alphabets of finite size can lead to improved performance. There are numerous different approaches towards the design of such decoders.  
For example, the authors of \cite{planjery2013a}, \cite{declercq2013a} and \cite{cai2014a} consider \ac{LUT} based update rules that are designed
such that the resulting decoders can correct most of the error events contributing to the error floor. 
However, their design is restricted to codes with column weight $3$ and to binary output channels. 
In~\cite{zhang2014a} a quasi-uniform quantization was proposed which extends the dynamic range of the messages at later iterations 
and improves the error floor performance. However, the design of~\cite{zhang2014a} still relies on the conventional message update 
rules and therefore does not reduce the required message bit-width. Finally, the authors of \cite{kurkoski2008a,kurkoski2014a} consider 
message updates based on an information theoretic fidelity criterion. While
\cite{planjery2013a},
\cite{declercq2013a}, and 
\cite{zhang2014a}
analyze the performance of their decoding schemes by means of \ac{FER} simulations, 
\cite{kurkoski2008a}
only provides density evolution results and
\cite{kurkoski2014a}
focuses solely on the algorithm for designing the message update rules. To the best of our knowledge,
none of the above schemes have been assessed in terms of their impact on hardware implementations.

\subsubsection*{Contribution}
In this paper, we derive a low-complexity decoding algorithm that is designed to directly operate with a finite message alphabet and that manages to achieve better error-rate performance than conventional algorithms with message resolutions as low as $3$ bits. 
Based on this algorithm, we synthesize a fully unrolled \ac{LDPC} decoder and compare our results
with our implementation of the only existing fully unrolled \ac{LDPC} decoder~\cite{schlafer2013a}. Our approach for 
the design of the \acl{VN} update rule is similar to \cite{kurkoski2008a,kurkoski2014a}, but we use a more sophisticated tree structure 
as well as a different \acl{CN} update rule.

\section{\acs{LDPC} Codes and Min-Sum Decoding}\label{sec:ldpc}

A $(d_v,d_c)$-regular \ac{LDPC} code is the set of codewords
\begin{equation}
   \big\lbrace \vec c \in \{0,1\}^N  \big | \mat H \vec c = \vec 0  \big\rbrace,
\end{equation}
where all operations are performed modulo~$2$. The parity check matrix $\mat H \in \{0,1\}^{M\times N}$ 
contains $d_v$ ones per column and $d_c$ ones per row and is sparse in the sense that $d_v < d_c \ll N$. 
The parity-check matrix forms an incidence matrix for a Tanner graph which contains $N$ 
\acp{VN} and $M$ \acp{CN}. Variable node $n$ is connected to check node $m$ if and only if $\mat H _{mn} = 1$.

\ac{LDPC} codes are traditionally decoded using \ac{MP} algorithms, where information is exchanged between the
\acp{VN} and the \acp{CN} over the course of several decoding iterations. Let the message alphabet be denoted by $\mathcal{M}$. 
For simplicity, in this work we assume that $\set M$ does not change over the iterations. At each iteration the messages from 
\acs{VN} $n$ to \acs{CN} $m$ are computed using the mapping 
$\Phi_v: \set L \times \set M^{d_v-1}\rightarrow \set M$, which is defined as
\begin{align} \label{eqn:vnupdateGeneric}
        \mu_{n\rightarrow m} = \Phi_v \big(L_n, \vec{\bar\mu}_{{\mathcal N}(n)\setminus m \rightarrow n} \big),
\end{align}
where $\set N(n)$ denotes the neighbours of node $n$ in the Tanner graph, $\vec{\bar\mu}_{{\mathcal N}(n)\setminus m \rightarrow n} \in \set M^{d_v-1},$ 
is a vector that contains the incoming messages from all neighboring \acp{CN} except $m$, and $L_n \in \set L$ denotes the channel \ac{LLR}  
corresponding to \ac{VN} $n$.
Similarly, the \acs{CN}-to-\acs{VN} messages are computed using the mapping $\Phi_c: \set M^{d_c-1}\rightarrow \set M$, which is defined as
\begin{equation} \label{eqn:cnupdateGeneric}
    \bar\mu_{m\rightarrow n} = \Phi_c \big(\vec{\mu}_{    {\mathcal N}(m)\setminus n \rightarrow m} \big),
\end{equation}
\Cref{fig:message_updates} illustrates the message updates in the Tanner graph. 
In addition to $\Phi_v$ and $\Phi_c$, a third mapping $\Phi_d: \set L \times \set M^{d_v}\rightarrow \{0,1\}$ 
is needed to provide an estimate of the transmitted codeword bit based on the incoming 
check node messages and the channel \ac{LLR} $L_n$
\begin{equation}\label{eqn:decupdate}
   \hat c_n = \Phi_d (L_n, \vec{\bar \mu}_{\set N(n)\rightarrow n}).
\end{equation}
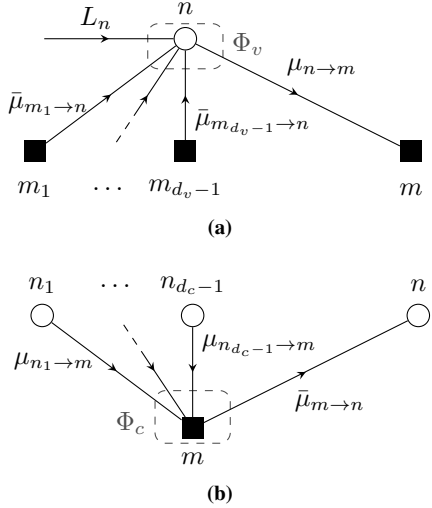
\begin{figure}%
   \centering
      \subfloat[][]{ 
\begin{tikzpicture}[
   >=stealth,
   decoration={
    markings,
    mark=at position 0.5 with {\arrow{>}}}
    ] 

\tikzstyle{cnode}=[rectangle, inner sep = 4pt, fill=black]
\tikzstyle{vnode}=[draw=black, circle, inner sep =3pt]
   \draw (0   , 0) node[cnode] (m1) {};
   \draw (0   , -.5) node{$m_1$};

   \node (mphantom)  at (1   , 0) {};
   \draw (1   , -.5) node{$\dots$};

   \draw (2   , 0) node[cnode] (md) {};
   \draw (2   , -.5) node{$m_{d_v-1}$};

   \draw (5   , 0) node[cnode] (m) {};
   \draw (5   , -.5) node{$m$};

   \draw (2   , 1.5)   node[vnode] (n) {};
   \draw (2   , 1.9)  node{$n$};
   \draw[dashed, rounded corners=4pt, black!70]   (1.5   , 1.1) rectangle (2.5, 1.7) node[below right] {$\Phi_v$};

   \node (llr)  at (0   , 1.5) {};

   \draw[postaction={decorate}] (m1) -- (n) node[pos =.4, left]{$\bar\mu_{m_1\rightarrow n}$};
   \draw[postaction={decorate}, dash pattern=on 3pt off 3pt on 3pt off 3pt on 500pt] (mphantom) -- (n);
   \draw[postaction={decorate}] (md) -- (n) node[pos =.2, right]{$\bar\mu_{m_{d_v-1}\rightarrow n}$};
   \draw[postaction={decorate}] (n) -- (m) node[pos =.4, above right]{$\mu_{n\rightarrow m}$};

   \draw[postaction={decorate}] (llr) -- (n) node[pos =.4, above]{$L_n$};

\end{tikzpicture}
 \label{subfig:vnupdate} } \\
      \subfloat[][]{ 
\begin{tikzpicture}[
   >=stealth,
decoration={
    markings,
    mark=at position 0.5 with {\arrow{>}}}
    ] 

\tikzstyle{cnode}=[rectangle, inner sep = 4pt, fill=black]
\tikzstyle{vnode}=[draw=black, circle, inner sep =3pt]
   \draw (0   , 1.5) node[vnode] (n1) {};
   \draw (0   , 1.9) node{$n_1$};

   \node (nphantom)  at (1   , 1.5) {};
   \draw (1   , 1.9) node{$\dots$};

   \draw (2   , 1.5) node[vnode] (nd) {};
   \draw (2   , 1.9) node{$n_{d_c-1}$};

   \draw (5   , 1.5) node[vnode] (n) {};
   \draw (5   , 1.9) node{$n$};

   \draw (2   , 0)   node[cnode] (m) {};
   \draw (2   , -.4)  node{$m$};
   \draw[dashed, rounded corners=4pt, black!70]   (2.5   , .5 ) rectangle (1.5, -.2) node[above left] {$\Phi_c$};

   \draw[postaction={decorate}] (n1) -- (m) node[pos =.4, left]{$\mu_{n_1\rightarrow m}$};
   \draw[postaction={decorate}, dash pattern=on 3pt off 3pt on 3pt off 3pt on 500pt] (nphantom) -- (m);
   \draw[postaction={decorate}] (nd) -- (m) node[pos =.2, right]{$\mu_{n_{d_c-1}\rightarrow m}$};
   \draw[postaction={decorate}] (m) -- (n) node[pos =.4, below right]{$\bar\mu_{m\rightarrow n}$};

\end{tikzpicture}
 \label{subfig:cnupdate} }
      \caption{
      \acs{VN} update 
      \protect\subref{subfig:vnupdate}  
      and \acs{CN} update 
      \protect\subref{subfig:cnupdate} 
      for $\set N (n) =\{m,m_1,\dots, m_{d_{v}-1}\}$ and $\set N (m) =\{n,n_1,\dots, n_{d_{c}-1}\}$ 
      }%
      \label{fig:message_updates}%
\end{figure}
For the widely used \ac{MS} algorithm, the mappings read
\begin{align}
   \label{eqn:vnupdateMS}
   &\Phi_v^{\mathrm{MS}}  (L, \vec{\bar\mu}  \big) = L + \sum_i \bar\mu_i, \\ 
   \label{eqn:cnupdateMS}
   & \Phi_c^{\mathrm{MS}} (\vec \mu \big)
        = \sign \vec \mu \,\min |\vec \mu| ,
\end{align}
where $\min |\vec \mu|$ denotes the minimum of the absolute values of the vector components, $\sign \vec \mu = \prod_j \sign \mu_j$
and $\set M = \set L = \mathbb R$. 
The decision mapping $\Phi_d$ is defined as
\begin{equation} \label{eqn:minsumDec}
   \Phi_d^{\mathrm{MS}}(L,\vec{\bar\mu}) =  \frac{1}{2}\left(1-\sign \left (L + \sum_i \bar\mu_i \right)\right).
\end{equation}

\section{Finite-Alphabet Decoder Design algorithm}\label{sec:decoder_design}

The MS algorithm assumes that the message set $\set M$ and the \ac{LLR} set $\set L$ are real numbers. However, it is 
impractical to use floating-point arithmetic in hardware implementations of such decoders and the message alphabets are 
usually discretized using a relatively low number of uniformly spaced quantization levels. This uniform quantization,  
together with the well-established two's complement and sign-magnitude binary encoding,  
leads to efficient arithmetic circuits, but it is not necessarily the best choice in terms of error-rate performance. 

Recently, efforts have been made to devise decoders that are designed to work directly with finite message and \ac{LLR} 
alphabets~\cite{planjery2013a,kurkoski2008a}. Instead of arithmetic computations such as \cref{eqn:vnupdateMS} and 
\cref{eqn:cnupdateMS}, the update rules for these decoders are implemented as look-up tables (LUTs). 
There are numerous approaches to the design of such LUTs. In the following, we provide an algorithm that is 
a mixture between the conventional \ac{MS} algorithm and purely \ac{LUT}-based decoders. More specifically, 
we only replace the \acs{VN} update rules with \acp{LUT}, which are designed using an information theoretic metric.  
For the design of the \acp{CN}, we exploit the fact that the outputs of the \ac{LUT}-based \acp{VN}, although not representing real numbers, can be ordered and for symmetric channels,
the message sign can be directly inferred from the labels, cf. \cref{subsec:QuantDiscussion}. This allows us to use the standard \ac{MS} update rule, thereby avoiding the
high hardware complexity that a \ac{LUT}-based \ac{CN} design would cause for codes with high \ac{CN} degree. 
Our hybrid algorithm provides excellent performance even with very few message levels and leads to an 
efficient hardware architecture, which is described in detail in Section~\ref{sec:architecture}.

\subsection{Mutual Information Based \acs{VN} \ac{LUT} Design} \label{subsec:IT}
The key idea behind the LUT design method that we employ is that, given the CN-to-VN message distributions 
of the previous iterations, one can design the VN LUTs for each iteration in a way that maximizes the mutual information 
between the VN output messages and the codeword bit corresponding to the VN in question.

We first describe how the distribution of the CN-to-VN messages can be computed based on the distribution of the incoming 
CN-to-VN messages. If the Tanner graph is cycle-free, then the individual input messages of a \ac{CN} at iteration
$i$ are iid conditioned on the transmitted bit $\rv x$\footnote{In the following, random variables are denoted by sans-serif letters.}, and their distribution is denoted by $p^{(i)}_{\rv m | \rv x} (\mu | x)$. 
Then, the joint distribution of the $(d_c-1)$ incident messages 
conditioned on the transmitted bit value corresponding to the recipient \ac{VN} 
(cf. Fig.~\ref{fig:message_updates}) reads
\begin{equation}\label{eqn:CNprodChannel}
   p^{(i)}_{\vec{{\rv m}} | \rv x} (\vec\mu | x) =
     \left(\frac{1}{2}\right)^{dc-2} 
   \sum_{\vec x:\, \bigoplus \vec x = x}
   \prod_{j=1}^{d_c-1}
      p^{(i)}_{\rv m | \rv x} (\mu_j | x_j),
\end{equation}
where $\bigoplus \vec x$ denotes the modulo-$2$ sum of the components of $\vec x$.
Using the update rule \cref{eqn:cnupdateMS}, the distribution of the outgoing
\acs{CN}-to-\acs{VN} message is then given by
\begin{equation} \label{eqn:CNmsgDist}
   p^{(i)}_{\overline{\rv m} | \rv x}(\bar\mu | x) =
   \sum_{\vec \mu \in \set M_{\bar\mu}} 
    p^{(i)}_{\vec{{\rv m}} | \rv x} (\vec\mu | x_n),
\end{equation}
where $\set M_{\bar\mu} \defas \Big\{\vec \mu \;\,\Big|\;\,\min |\vec \mu| = |\bar\mu| \;\wedge\;\sign \vec \mu = \sign \bar\mu\Big\}$.  
The output message values are given by 
\begin{equation}\label{eqn:msgVal}
   \mu \defas \log \frac{p^{(i)}_{\rv m | \rv x}(\mu |0)}{p^{(i)}_{\rv m | \rv x}(\mu |1)}.
\end{equation}
Conventional decoding algorithms need a high dynamic range in order to represent the growing message magnitudes, as they are using the same message representation for every iteration. 
In our LUT-based decoder, the message representation changes from one iteration to the next and the message values grow implicitly as the distributions 
$p^{(i)}_{\rv m | \rv x}(\mu|x)$ become more and more concentrated over the course of the iterations, 
thus providing an explanation for the good performance we can achieve with very low resolutions.

Let $\vec{\bar\mu} = \begin{pmatrix} \bar\mu_1, \dots, \bar\mu_{d_v-1}\end{pmatrix}$ 
denote the $(d_v-1)$ incident \ac{CN}-to-\ac{VN} messages that are involved in the update of a certain \ac{VN} 
(one of which is always the channel LLR $L$) and let $\rv x$ be the transmit bit corresponding to this \ac{VN}. 
Then, the joint distribution of the VN input messages is given by~\cite{kurkoski2008a}
\begin{equation} \label{eqn:VNprodChannel}
       p^{(i)}_{\rv L,\vec{\overline{\rv m}} | \rv x} (L,\vec{\bar\mu} | x) =
   \hspace{-15 pt}\sum_{\vec x:\, x_0=\dots= x_{d_v-1} = x}\hspace{-15pt}
   p_{\rv L| \rv x} ( L | x_0 )
   \prod_{j=1}^{d_v-1} 
   p^{(i)}_{\overline{\rv m} | \rv x}(\bar\mu_j | x_j).
\end{equation}
Given this distribution, we can construct an update rule 
\begin{equation}\label{eqn:MIquantizer} 
   \Phi_{v}^{(i)\,\mathrm{MI}} = \argmax_{Q\in \set Q} I(\rv m; \rv x) = 
   \argmax_{Q\in \set Q} I\big(Q( \rv L, \overline{\rvec m}); \rv x\big),
\end{equation}
where $\set Q$ is the set of all deterministic mappings in the form of \cref{eqn:vnupdateGeneric} and $I(\rv m; \rv x)$ 
denotes the mutual information between $\rv m$ and $\rv x$. Hence, the resulting update rule locally 
maximizes the information flow between the \acp{CN} and the \acp{VN}.

An algorithm that solves \cref{eqn:MIquantizer} with complexity $O\big( |\set L|^3|\set M|^{3(d_v-1)}\big)$  was provided in~\cite{kurkoski2014a}.
Using the update rule \cref{eqn:MIquantizer}, we can compute the
message conditional distribution of the next iteration
\begin{equation}\label{eqn:VNmsgDist}
    p^{(i+1)}_{\rv m | \rv x}(\mu | x) =
   \hspace{-15 pt}\sum_{(L,\vec{\bar\mu)}:\,\Phi_{v}^{(i)\,\mathrm{MI}}(L,\vec{\bar\mu}) = \mu} \hspace{-15 pt}
   p^{(i)}_{\rv L,\overline{ \vec{\rv m}} | \rv x} (L,\vec{\bar\mu} | x).
\end{equation}

Given an initial message distribution 
$p^{(0)}_{\rv m | \rv x}(\mu | x) $
and a distribution of the channel \acp{LLR} 
$p_{\rv L | \rv x}(L | x) $, the repeated alternating application of
\cref{eqn:CNprodChannel,eqn:CNmsgDist,eqn:VNprodChannel,eqn:MIquantizer,eqn:VNmsgDist}
produces a sequence of locally optimal \ac{VN} update mappings 
$\Phi_{v}^{(i)\,\mathrm{MI}},~i \in \{1,\hdots,I\}$, where $I$ denotes a pre-determined maximum 
number of performed iterations.

\subsection{Discussion and Practical Considerations}\label{subsec:QuantDiscussion}

\subsubsection{\acs{LUT}-based \acs{VN} and Tree Structure}
As the mappings $\Phi_v$  take $|\set L|\cdot|\set M|^{(d_v-1)}$ inputs, a direct application of
the algorithm described in Section~\ref{subsec:IT} is restricted to low weight codes. However, we can
construct a hierarchy of mappings where each partial mapping only processes a subset of the inputs 
and the intermediate outputs of preceding stages.\footnote{In our simulations, we observe 
that the choice of LUT tree structure can significantly affect the FER performance of the decoder. 
It is an interesting open problem to identify the best possible tree structure under some complexity 
constraint (e.g., we could limit the number of LUT inputs to two).} The quantizer design for such a hierarchy follows 
directly by considering only the messages incident to the respective mapping in  
\cref{eqn:VNprodChannel} and, for the intermediate nodes, replacing the distributions \cref{eqn:CNmsgDist} 
of the incident \ac{CN}  messages by the distributions \cref{eqn:VNmsgDist} of the previous stage.

\subsubsection{Channel Output Quantization}
So far we considered the initial distributions 
$p^{(0)}_{\rv m | \rv x}(\mu | x) $ 
and
$p_{\rv L | \rv x}(L | x) $
as given. When designing practical decoders for communication applications, the initial distributions
follow from the transmission channel and the \ac{LLR} quantization of the preceding signal processing. 

Throughout the rest of the paper, we consider a \ac{BI-AWGN} channel followed by maximum mutual information 
quantization of the \acp{LLR} \cite{winkelbauer2015a}. In this case, the initial distributions depend 
on the SNR, which renders the LUT design SNR-specific. Nevertheless, we observe in our simulations that 
the decoder generally performs well also for SNRs other than the design SNR.

\subsubsection{Message Representation for Symmetric Channels}
Consider the practically relevant case where $|\set M|$ and $|\set L|$ are even and the distributions 
$p^{(0)}_{\rv m | \rv x}(\mu | x)$ 
and
$p_{\rv L | \rv x}(L | x)$
are symmetric in the sense that
\begin{align}
p_{\rv L | \rv x}(L_k|0) &\equiv  p_{\rv L | \rv x}(L_{|\set L|-k+1}|1), & &k=1,\dots,\frac{|\set L|}{2}\\
p^{(0)}_{\rv m | \rv x}(\mu_k|0) &\equiv  p^{(0)}_{\rv m | \rv x}(\mu_{|\set M|-k+1}|1), & &k=1,\dots,\frac{|\set M|}{2}
\end{align}
or equivalently, expressed in terms of the \acp{LLR} values
\begin{equation}\label{eqn:llr_symmetry_disc}
   L_k  \equiv -L_{|\set L|-k+1}  \qquad  \mu_k  \equiv -\mu_{|\set M|-k+1}.
\end{equation}
For that case, computing the \ac{CN} update \cref{eqn:cnupdateMS} is simplified as the
sign follows immediately from the message labels. Thus, for the \ac{CN} update the message 
values do not need to be stored and the entire decoder can be implemented based on the message labels.

\subsubsection{Decision Stage}\label{sec:decstage}
Since the discrete messages of our decoder do not represent real numbers but are labels, a simple arithmetic decision 
mapping such as \cref{eqn:minsumDec} is not possible. Instead, $\Phi_d$ has to be implemented as a generic
mapping as well. The construction of  $\Phi_d$  is similar to the construction of $\Phi_v$, with the 
difference that all $d_v$ input messages and the channel \ac{LLR} have to be processed and that the output is binary.

\begin{figure*}
  \centering
  \includegraphics[width=1\textwidth]{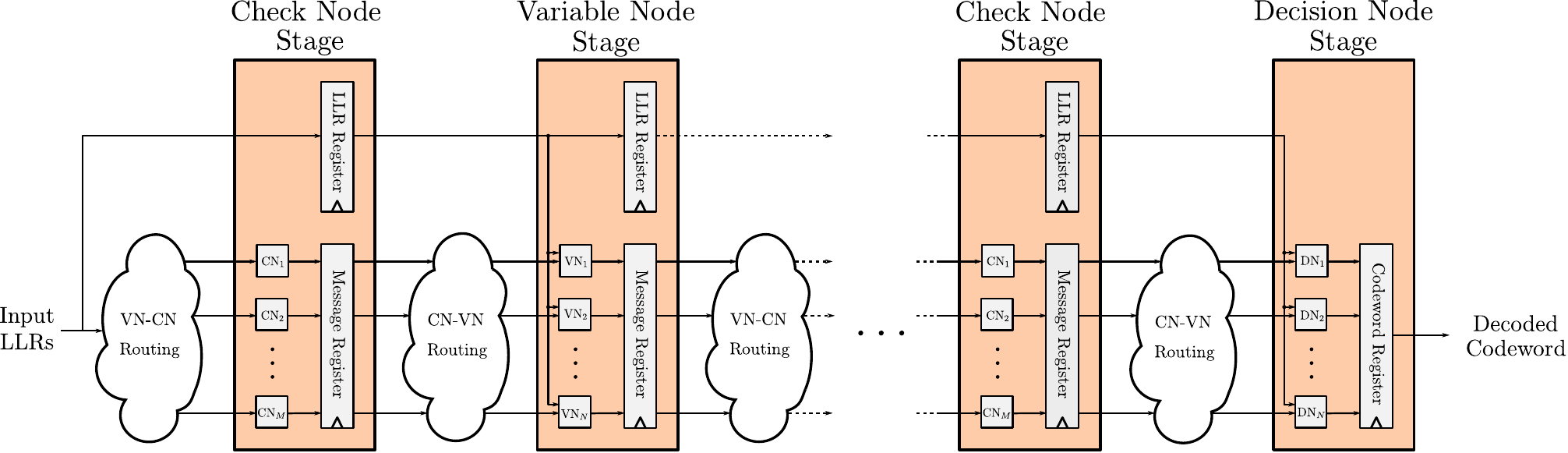}
	\caption{Top level decoder architecture processing pipeline. The channel LLRs are the input of the left-hand side and the decoded codeword is obtained as the output of the right-hand side.}\label{fig:toplevel}
	\vspace{-0.15cm}
\end{figure*}

\section{LUT-Based Fully Unrolled Decoder Hardware Architecture}\label{sec:architecture}
In the previous section, we have described an algorithm that can construct locally optimal variable node update rules in the form of \acp{LUT} for a given quantization bit-width for each iteration for any given $(d_v,d_c)$-regular LDPC code. Most conventional LDPC decoder architectures are either partially parallel, meaning that fewer than $N$ VNs and $M$ CNs are instantiated, or fully parallel, meaning that $N$ VNs and $M$ CNs are instantiated. Using a LUT-based decoder with a carefully designed quantization scheme can significantly reduce the memory required to store the messages exchanged by the VNs and CNs due to the reduced message bit-width required to achieve the same FER performance. However, both for partially parallel and for fully parallel decoders, separate LUTs would be required within each VN for each one of the performed decoding iterations, significantly increasing the size of each VN, and thus possibly outweighing the gain in the memory area.

An additional degree of parallelism was recently explored in~\cite{schlafer2013a}, where a \emph{fully unrolled} and fully parallel LDPC decoder was presented. This decoder instantiates $N$ VNs and $M$ CNs for each iteration of the decoding algorithm, leading to a total of $N I$ VNs and $M I$ CNs. While such a fully unrolled decoder requires significant hardware resources, it also has a very high throughput since one decoded codeword can be output in each clock cycle. Thus, the hardware efficiency (i.e., throughput per unit area) of the fully unrolled decoder presented in~\cite{schlafer2013a} turns out to be significantly better than the hardware efficiency of partially parallel and fully parallel (non-unrolled) approaches. Since in a fully unrolled LDPC decoder architecture VNs and CNs are instantiated for each iteration, it is a very suitable candidate for the application of our LUT-based decoding algorithm.

In this section, we describe the hardware architecture of our fully unrolled LUT-based LDPC decoder. Our hardware architecture is similar to the architecture used in~\cite{schlafer2013a}, while the most important differences are the optimized LUT-based variable node and the significantly reduced bit-width of all quantities involved in the decoding process. 

\subsection{Decoder Architecture}
An overview of our decoder architecture is shown in Fig.~\ref{fig:toplevel}. Each decoding iteration is mapped to a distinct set of variable nodes and check nodes which then form a processing pipeline. In essence, a fully unrolled and fully parallel LDPC decoder is a systolic array in which data flows from left to right. A new set of $N$ channel LLRs can be read in each clock cycle, and a new decoded codeword is output in each clock cycle. The decoding latency as well as the maximum frequency depend on the number of performed iterations as well as the number of pipeline registers present in the decoder. Our decoder consist of three types of stages, namely the CN stage, the VN stage, and the DN stage, which are described in detail in the sequel. As long as a steady flow of input channel LLRs can be provided to the decoder, there is no control logic required apart from the clock and reset signals.

\subsubsection{Check Node Stage}
Each CN stage contains $M$ check node units, as well as $Md_c$ $Q_{\text{msg}}$-bit registers which store the check node output messages, where $Q_{\text{msg}}$ denotes the number of bits used to represent the internal decoder messages. Moreover, each CN stage contains $N$ $Q_{\text{ch}}$-bit channel LLR registers which are used to forward the channel LLRs required by the following variable node stages, where $Q_{\text{ch}}$ denotes the number of bits used to represent the channel LLRs.

Due to \eqref{eqn:llr_symmetry_disc}, we can use a check node architecture which is practically identical to the check node architecture used in~\cite{schlafer2013a}. More specifically, each check node consists of a sorting unit that identifies the two smallest messages among all $d_c$ input messages and an output unit which selects the first or the second minimum for each output, along with the appropriate sign. The sorting unit contains 4-input compare-and-select (CS) units in a tree structure, which identify and output the two smallest values out of the four input values~\cite{schlafer2013a}. We use sign-magnitude (SM) to represent all message labels. The SM2TC unit used in the check node of~\cite{schlafer2013a} is not required in our architecture since the variable node does not perform any arithmetic operations where the two's complement representation could be favorable.

\subsubsection{Variable Node Stage}
Each \ac{VN} stage contains $N$ variable node units, as well as $Nd_v$ $Q_{\text{msg}}$-bit registers that store the variable node output messages. Moreover, each VN stage contains $N$ $Q_{\text{ch}}$-bit channel LLR registers which are used to forward the channel LLRs required by the following VN stages.

In the variable node architecture used in the adder-based decoder of~\cite{schlafer2013a}, all input messages are added and then the input message corresponding to each output is subtracted from the sum in order to form the output message, thus implementing the conventional MS update rule given in~\eqref{eqn:vnupdateMS}. In order to avoid overflows, in our implementation of~\cite{schlafer2013a} the bit-width of the internal signals is increased by one bit for each addition.

\begin{figure}[t]
  \centering
  \subfloat[][]{ 
\tikzset{
leavenode/.style = {align=center, 
                    inner sep=2pt, 
                    text centered
                   },
imnode/.style = {align=center, 
                 inner sep=2pt, 
                 text centered, 
                 draw,
                 rounded corners = 1pt
                }
}
\begin{tikzpicture}
[<-,
>=stealth,
level 1/.style={sibling distance=1.4cm},
level 2/.style={sibling distance=1.4cm}, 
level 3/.style={sibling distance=1cm},
level 4/.style={sibling distance=.6cm},
level distance = .7cm
] 
        \node (root)[imnode] {LUT}
        child{ node [imnode] {LUT}
           child{ node [imnode] {LUT}
              child{ node [imnode] {LUT}
                 child{ node [leavenode] {$\bar\mu$}
                 }
                 child{ node [leavenode] {$\bar\mu$}
                 }
              }
              child{ node [imnode] {LUT}
                 child{ node [leavenode] {$\bar\mu$}
                 }
                 child{ node [leavenode] {$\bar\mu$}
                 }
              }
           }
           child{ node [imnode] {}  
              child{ node [imnode] {}
                 child{ node [leavenode] {$\bar\mu$}
                 }
              }
            } 
        }
        child{ node [imnode] {}
           child{ node [imnode] {}  
              child{ node [imnode] {}
                 child{ node [leavenode] {$L$}
                 }
              }
            } 
        } 
        ; 
        \draw[draw=black, ->] (root) --++ (1, 0) node[right]{$\mu$};
\end{tikzpicture}
 \label{fig:vnodelut} } \vspace{.7cm}
  \subfloat[][]{ 
\tikzset{
leavenode/.style = {align=center, 
                    inner sep=2pt, 
                    text centered
                   },
imnode/.style = {align=center, 
                 inner sep=2pt, 
                 text centered, 
                 draw,
                 rounded corners = 1pt
                }
}
\begin{tikzpicture}
[<-,
>=stealth,
level 1/.style={sibling distance=1.2cm},
level 2/.style={sibling distance=.4cm}, 
level distance = .7cm
] 
        \node (root) [imnode] {LUT} 
        child{ node [imnode] {LUT}
           child{ node [leavenode]{$\bar\mu$}
           }
           child{ node [leavenode]{$\bar\mu$}
           }
           child{ node [leavenode]{$\bar\mu$}
           }
        }
        child{ node [imnode] {LUT}
           child{ node [leavenode]{$\bar\mu$}
           }
           child{ node [leavenode]{$\bar\mu$}
           }
           child{ node [leavenode]{$\bar\mu$}
           }
        }
        child{node [imnode]{}
           child{node [leavenode]{$L$}
           }
        }
        ;
        \draw[draw=black, ->] (root) --++ (1, 0) node[right]{$\hat c$}; 
\end{tikzpicture}
 \label{fig:decnodelut} } 
    \caption{ (a)
    A variable node LUT tree for the calculation of one output of a variable node of degree $d_v = 6$. Each LUT-based variable node contains $d_v$ such LUT trees, one for each of the ${d_v \choose d_v-1}$ possible combinations of input messages.\\
   (b) A decision node LUT tree for a variable node of degree $d_v = 6$. Each LUT-based decision node contains a single decision tree.}
	\vspace{-0.2cm}
\end{figure}
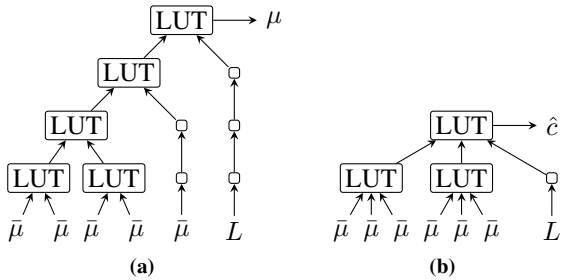

For our LUT-based decoder the adder tree is replaced by $d_v$ LUT trees, each of which computes one of the $d_v$ outputs of the variable node. One possible LUT-tree structure is shown in Fig.~\ref{fig:vnodelut}, where $\bar\mu$ denotes an internal message from a check node and $L$ denotes the channel LLR. LUT sharing between the $d_v$ LUT trees can be achieved by identifying the nodes that appear in more than one tree and instantiating them only once, thus significantly reducing the required hardware resources. Moreover, keeping the number of inputs of each LUT as low as possible ensures that the size of the LUTs, which grows exponentially with the number of inputs, is manageable for the automated logic synthesis process.

\subsubsection{Decision Node Stage}
The variable node that corresponds to the final decoding iteration is called a \emph{decision node} (DN). The DN stage contains $N$ decision nodes, as well $N$ single-bit registers that store the decoded codeword bits. The DN stage does not contain channel LLR registers, as there are no subsequent decoding stages where the channel LLRs would be used. 
The architecture of a decision node is generally simpler than that of a variable node, as a single output value (i.e., the decoded bit) is calculated instead of $d_v$ distinct outputs.

More specifically, in the architecture of~\cite{schlafer2013a}, the decision metric of~\eqref{eqn:decupdate} is already calculated as part of the variable node update rule. However, for the decision node, there is no need to subtract each input message from the sum in order to generate $d_v$ distinct output messages. It suffices to check whether the sum is positive or negative, and output the corresponding decoded codeword bit. 

In our LUT-based decoder, as discussed in Section~\ref{sec:decstage}, a LUT tree is designed whose tree node has an output bit-width of a single bit, which is the corresponding decoded codeword bit. An example of a decision LUT tree for a decision node that corresponds to a code with $d_v = 6$ is shown in Fig.~\ref{fig:decnodelut}. Each decision node contains a single LUT tree, in contrast with the variable nodes which contain $d_v$ LUT trees.

\subsection{Decoding Latency and Throughput}
Our LUT-based architecture contains pipeline registers at the output of each stage (VN, CN, and DN). Thus, for a given number of decoding iterations $I$, the decoding latency is $2I$ clock cycles. Since one decoded codeword is output in each clock cycle, the decoding throughput of the decoder, measured in Gbits/s, is given by
$T = Nf$, 
where $f$ denotes the operating frequency measured in GHz.

\subsection{Memory Requirements}
Each pipeline stage except the DN stage requires an $N Q_{\text{ch}}$ channel LLR register. Moreover, each VN and CN stage requires $N d_vQ_{\text{msg}}$ (equivalently, $M d_cQ_{\text{msg}}$) registers to store the output messages. Finally, the DN stage requires $N$ registers to store the decoded codeword bits. Thus, the total number of register bits required by our LUT-based decoder can be calculated as
\begin{align}
	B_{\text{tot}}	& = (2I-1)N (d_v Q_{\text{msg}} + Q_{\text{ch}}) + N. \label{eqn:totbits}
\end{align}
Naturally, \eqref{eqn:totbits} can also be used to calculate the register bits required by an adder-based MS architecture with the same pipeline register structure.

\section{Implementation Results}
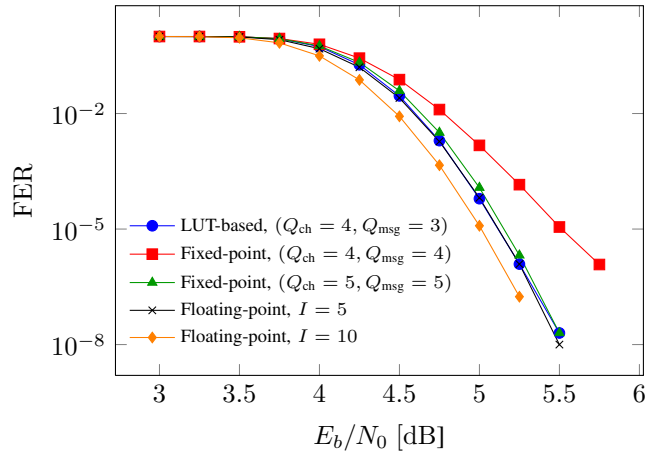
\begin{figure}[t]
   \centering
\begin{tikzpicture}
\begin{semilogyaxis}[
    width=8.6cm,
    height=6.5cm,
    xlabel={$E_b/N_0\;[\mathrm{dB}$]},
    ylabel={$\mathrm{FER}$},
    title={},
    legend style={
        cells={anchor=west}, 
        draw = none,         
        fill = none,         
        at={(.01,.05)},        
        anchor=south west    
    }
   ]

   \addplot+[
      mark=*,
      mark options={fill=blue}, 
      color=blue,
      ] 
      table [x index             = 0,
             y  index            = 1, 
             col sep=comma] {lut.csv};
   \addlegendentry{\scriptsize LUT-based, $(Q_{\text{ch}} = 4, Q_{\text{msg}} = 3)$}

   \addplot+[
      mark=square*,
      mark options={fill=red},
      color = red,
      ] 
      table [x index             = 0,
             y  index            = 1, 
             col sep=comma] {fixed_4bit.csv};
   \addlegendentry{\scriptsize Fixed-point, $(Q_{\text{ch}} = 4, Q_{\text{msg}} = 4)$}

   \addplot+[
      mark=triangle*, 
      mark options={fill=green!60!black},
      color = green!60!black,
      ] 
      table [x index             = 0,
             y  index            = 1, 
             col sep=comma] {fixed_5bit.csv};
   \addlegendentry{\scriptsize Fixed-point, $(Q_{\text{ch}} = 5, Q_{\text{msg}} = 5)$}

   \addplot+[
      mark=x,
      mark options={fill=black},
      color = black,
      ] 
      table [x index             = 0,
             y  index            = 1, 
             col sep=comma] {float_05iter.csv};
   \addlegendentry{\scriptsize Floating-point, $I=5$}

   \addplot+[
      mark=diamond*,
      mark options={fill=red!50!yellow},
      color = red!50!yellow,
      ] 
      table [x index             = 0,
             y  index            = 1, 
             col sep=comma] {float_10iter.csv};
   \addlegendentry{\scriptsize Floating-point, $I=10$}

\end{semilogyaxis}
\end{tikzpicture}

   \caption{FER vs $E_b/N_0$ for the $N=2048$ $(6,32)$-regular LDPC code defined in IEEE 802.3an.}\label{fig:fer} \vspace{-.6cm}
\end{figure}

In this section, we present synthesis results for a fully unrolled LUT-based LDPC decoder and we compare it with synthesis results of our implementation of a fully unrolled adder-based MS  LDPC decoder. We have used the parity-check matrix of the LDPC code defined in the IEEE 802.3an standard~\cite{802.3an} ($10$ Gbit/s Ethernet), which is a $(6,32)$-regular LDPC code of rate $R = 13/16$ and blocklength $N = 2048$. For the fixed point decoder and the LUT-based
 decoder, a total of $I = 5$ decoding iterations are performed, since from Fig.~\ref{fig:fer} we observe that increasing the number of iterations to, e.g., $I=10$, does not lead to a significant improvement in performance for this LDPC code. All synthesis results are obtained by using a TSMC~$90$nm CMOS library under typical operating conditions.

\subsection{Quantization Parameters}
For the LUT-based decoder, we have used $Q_{\text{ch}} = 4$ bits for the representation of the channel LLRs and $Q_{\text{msg}} = 3$ bits for the representation of the internal messages, as this leads to an error correction performance that is very close the floating-point MS decoder (cf. Fig.~\ref{fig:fer}). For the variable nodes, we use the LUT tree structure of Fig.~\ref{fig:vnodelut} and for the decision nodes we use the LUT tree structure of Fig.~\ref{fig:decnodelut}. The design SNR is set to $4.5$~dB. For the adder-based MS decoder which serves as a reference, we use $Q_{\text{ch}} = 5$ bits for the representation of the channel LLRs and $Q_{\text{msg}} = 5$ bits for the representation of the internal messages, as this leads to practically the same FER performance for the LUT-based and the adder-based MS decoder, as can be seen in Fig.~\ref{fig:fer}. 

\begin{table}[t]
\caption{Synthesis Results}\label{tab:results}
\centering
\begin{tabular}{c|cc}
	   							& Adder-based MS			& LUT-based \\
\hline
Area 							& $35.63$~mm$^2$			& $33.79$~mm$^2$ \\
Frequency					& $495$~MHz						& $813$~MHz \\
Latency						&	$20.20$~ns					& $12.30$~ns \\
Throughput				& $1014$~Gbps					& $1665$~Gbps \\
Area Efficiency 	& $28.46$~Gbps/mm$^2$	& $49.27$~Gbps/mm$^2$
\end{tabular}
\vspace{-0.0cm}
\end{table}

\begin{table}[t]
\caption{Area Breakdown}\label{tab:breakdown}
\centering
\begin{tabular}{c|rr}
	   								& Adder-based MS& LUT-based \\
\hline
										& \multicolumn{2}{c}{Check Node Stage} \\
\hline
Check Nodes					& $2.77$~mm$^2$ 				& $1.11$~mm$^2$ \\
Pipeline Registers	& $1.11$~mm$^2$ 				& $0.70$~mm$^2$ \\
Total								& $3.88$~mm$^2$					& $1.81$~mm$^2$ \\				
\hline
										& \multicolumn{2}{c}{Variable Node Stage} \\
\hline
Variable Nodes $I_1$& $2.35$~mm$^2$ 				& $4.62$~mm$^2$ \\
Variable Nodes $I_2$& $2.35$~mm$^2$ 				& $4.78$~mm$^2$ \\
Variable Nodes $I_3$& $2.35$~mm$^2$ 				& $4.64$~mm$^2$ \\
Variable Nodes $I_4$& $2.35$~mm$^2$ 				& $4.68$~mm$^2$ \\
Pipeline Registers	& $1.11$~mm$^2$ 				& $0.57$~mm$^2$ \\
Total	 $I_1$				& $3.46$~mm$^2$ 				& $5.32$~mm$^2$ \\	
Total	 $I_2$				& $3.46$~mm$^2$ 				& $5.48$~mm$^2$ \\	
Total	 $I_3$				& $3.46$~mm$^2$ 				& $5.34$~mm$^2$ \\
Total	 $I_4$				& $3.46$~mm$^2$ 				& $5.38$~mm$^2$ \\	
\hline
										& \multicolumn{2}{c}{Decision Node Stage}	\\
\hline
Decision Nodes			& $2.035$~mm$^2$				& $3.21$~mm$^2$ \\
Pipeline Registers	& $0.03$~mm$^2$				& $0.03$~mm$^2$ \\
Total								& $2.38$~mm$^2$				& $3.24$~mm$^2$ \\
\hline
										& \multicolumn{2}{c}{Top-Level Decoder} \\
\hline
Logic Area					& $25.58$~mm$^2$				& $27.46$~mm$^2$ \\
Register Area				& $10.05$~mm$^2$				& $6.33$~mm$^2$ \\
Total Area 					& $35.63$~mm$^2$			 	& $33.79$~mm$^2$ \\
\end{tabular}
\vspace{-0.3cm}
\end{table}

\subsection{Adder-based vs. LUT-based Decoder}
We present synthesis results for the adder-based and the LUT-based decoders in Table~\ref{tab:results}. For fair comparison, we synthesized both designs for various clock constraints and selected the result with the highest hardware efficiency for each design. These results should not be regarded in absolute terms, as the placement and routing of such a large design is highly non-trivial and will increase the area and the delay of both designs significantly. However, it is safe to make relative comparisons, especially when considering the fact that the LUT-based decoder will be easier to place and route due to the fact that it requires approximately 40\% fewer wires for the interconnect between the VN, CN, and DN stages. We observe that the LUT-based decoder is approximately 8\% smaller as well as 64\% faster than the adder-based MS decoder. As a result, the area efficiency of the LUT-based decoder is $73\%$ higher than that of the adder-based MS decoder. For both designs, the critical path goes through the CN, but in the LUT-based decoder the delay is smaller due to the reduced bit-width.

We show the area breakdown of the LUT-based and the adder-based decoders in Table~\ref{tab:breakdown}. We observe that the VN stage area of the LUT-based decoder varies significantly over the iterations, even though the LUT tree structures are identical. This is not unexpected, since the contents of the LUTs are different for different iterations and the resulting logic circuits can have very different complexities. Moreover, we see that the CN stage of the LUT-based decoder is approximately $53\%$ smaller than the CN stage of the adder-based decoder due to the bit-width reduction enabled by the optimized LUT design. The VN stage of the LUT-based decoder, on the other hand, is larger than the VN stage of the adder-based decoder. However, the reduction in the CN stage is larger than the increase in the VN stage, leading to an overall reduction in area. From Table~\ref{tab:breakdown} we can see that this reduction stems mainly from the reduced number of required registers, as the area occupied by the logic of each decoder is similar.

\section{Conclusion}
In this paper, we described a method that can be applied to design a discrete message-passing decoder for LDPC codes by replacing the standard VN update rules with locally optimal LUT-based update rules. Moreover, we presented a hardware architecture for a LUT-based fully unrolled LDPC decoder which can reduce the area and increase the operating frequency compared to a conventional adder-based MS decoder by $8\%$ and $64\%$, respectively, due to the significantly reduced bit-width required to achieve identical error correction performance.  Finally, the LUT-based decoder requires approximately $40$\% fewer wires, simplifying the routing step, which is a known problem in fully parallel architectures.

\bibliographystyle{IEEEtran}

\end{document}